\documentstyle[multicol,pre,aps,psfig,array]{revtex}

\newcommand{\be}{\begin{eqnarray}}
\newcommand{\ee}{\end{eqnarray}}

\begin{document}
\title{Front Propagation in Chaotic and Noisy Reaction-Diffusion
Systems:\\a Discrete-Time Map Approach}
\author{Alessandro Torcini$^{1,2}$, Angelo Vulpiani$^{1,3}$, 
and Andrea Rocco$^{1,3}$}
\address{$^1$Dipartimento di Fisica, Universit\`a di Roma ``La Sapienza''
P.le Aldo Moro 2, I-00185 Roma, Italy}
\address{$^2$ Istituto Nazionale di Fisica della Materia, UdR Firenze,
L.go E. Fermi, 3 - I-50125 Firenze, Italy}   
\address{$^3$Istituto Nazionale di Fisica della Materia, UdR Roma, 
P.le Aldo Moro 2, I-00185 Roma, Italy}
\date{\today}

\maketitle

\draft

\begin{abstract}
We study the front propagation in Reaction-Diffusion systems whose 
reaction dynamics exhibits an unstable fixed point and chaotic or noisy
behaviour. We have examined the influence of chaos and noise on 
the front propagation speed and on the wandering of the front
around its average position. Assuming that the reaction 
term acts periodically in an impulsive way, the dynamical evolution of 
the system can be written as the convolution between a spatial propagator and
a discrete-time map acting locally. This approach allows us to perform 
accurate numerical analysis. They reveal that in the pulled regime 
the front speed is basically determined by the shape of the map 
around the unstable fixed point, while its chaotic or noisy features 
play a marginal role. In contrast, in the pushed regime the presence 
of chaos or noise is more relevant. 
In particular the front speed decreases when the degree of chaoticity 
is increased, but it is not straightforward to derive 
a direct connection between the chaotic 
properties ({\em e.g.} the Lyapunov exponent) and the behaviour 
of the front. As for the fluctuations of the front position, 
we observe for the noisy maps that the associated mean square
displacement grows in time  as $t^{1/2}$ in the pushed case and 
as $t^{1/4}$ in the pulled one, in agreement with recent findings
obtained for continuous models with multiplicative noise.
Moreover we show that the same quantity saturates when a 
chaotic deterministic dynamics is considered for both pushed and
pulled regimes.
\end{abstract}


%

\begin{multicols}{2}

\section{Introduction}
\label{intro}

In the last years the study of front dynamics has gained more and
more relevance in many different fields, such as chemistry \cite{kapral},
biology \cite{biology}, combustion \cite{Peters}. 
In physics \cite{combustion} the problem of front propagation is
generally related to Reaction-Diffusion (RD) systems, 
and to the identification of the different dynamical regimes present 
in the model under study. 

In this case the focus is often on
the situation where RD fronts connect a stable state to an unstable
one.  Consider for instance the prototype equation for front propagation
\be
\partial_t u = D \partial_{xx} u + f(u) \quad,
\label{fkpp}
\ee
where $u=u(x,t)$ , $D$ is the diffusion coefficient and $f$ is
a continuous function with two fixed points, $u=1$, stable, and 
$u=0$, unstable. This equation is usually referred to as 
Fisher-Kolmogorov-Petrovsky-Piskunov (FKPP) Equation \cite{fish,kpp}.

It is well known that if one considers an initial condition 
$u(x,0)$, different from zero in a bounded spatial region, a front 
develops connecting the two fixed points. 
The asymptotic leading edge, {\em i.e.} the
semi-infinite region ahead of the front, has typically an exponential
shape of the type $\exp(-\mu x + \lambda t)$. 
In principle,
depending on the value of the spatial decay-rate $\mu$, the front
speed can take a continuous set of values, namely 
$v(\mu) =  \lambda/\mu$ \cite{wim1}. 

However,
for sufficiently localized initial conditions a unique speed $v_{\rm f}$ is
selected, which is always bounded 
in the range $[v_{\rm lin},v_{\rm max}]$, with
\be
v_{\rm lin} = \min_{\mu} v(\mu) = 2 \sqrt{D f^\prime(0)}
\label{vmin}
\ee
and
\be
v_{\rm max} = 2 \sqrt{D \max_{0 \le u \le 1} \, \frac{f(u)}{u}}\quad.
\label{vmax}
\ee

If the function $f(u)$ is concave, it is possible
to prove that the propagation speed of the front
coincides with the linear one, $v_{\rm f} \equiv v_{\rm lin}$
\cite{ara}. In literature this case is often referred to as {\em pulled}
dynamics. The nonlinearities present in the system 
are not dynamically relevant, and a linearization about the 
unstable state is enough to fully describe the propagation of the
front. As a result the front is indeed ``pulled'' by the 
spreading and growth of linear perturbations in the leading edge \cite{wim2}.
On the other hand, the function $f(u)$ being not concave
is a necessary condition to have $v_{\rm f} > v_{\rm lin}$ \cite{ara}. This in
turn corresponds to the {\em pushed} regime, where the nonlinearities 
are relevant to the dynamics and the front is ``pushed'' by its internal 
part \cite{wim2}.

Such a dynamical behaviour can also be studied when inhomogeneous 
environments are considered. A typical example is the 
photosensitive Belousov-Zhabotinsky chemical reaction. This system 
allows for an experimental realization of external noise, achieved via 
fluctuating illumination conditions \cite{irene}. The random 
environment can be mimicked by multiplicative noise terms 
in the corresponding RD equation -- see for example 
\cite{armeroprl,armeropre,rocco1,rocco2}. Under
the effect of noise, the motion of the front can be
decomposed into a global drift, characterized by a noise-renormalized 
speed, plus fluctuations, affecting the position of the front. Particularly 
the wandering of the front around its average position has been proven to be
diffusive in the pushed case \cite{armeropre} and subdiffusive 
in the pulled one \cite{rocco2}, with an associated mean square
displacement growing in time respectively like $t^{1/2}$ and $t^{1/4}$.

Another natural way for erratic behaviors to occur
in RD systems is via a chaotic underlying process.
Even though so far only a limited number of papers has been devoted to 
the influence of chaotic dynamics on front propagation \cite{pikov,storm},
chaos is extremely relevant in RD systems describing chemical reactions
\cite{kapral}, as well as in more general pattern 
forming systems \cite{cross}. Moreover, in the context
of spatially extended chaotic systems many concepts have been
borrowed from the study of front propagation into unstable
steady states to describe the spreading of information \cite{torc1}.

In this paper we aim at taking up this issue again, by studying 
the front propagation in systems where an unstable fixed point 
is present and in addition the reaction dynamics is chaotic (or noisy).
In the chaotic case the role of the stable fixed point in the usual 
FKPP problem is played by the chaotic phase. 
The most natural way of realizing this situation is to consider 
the case ${\bf u} \in R^d$ with $d \ge 3$ in the FKPP Equation, 
that is 
\be
\partial_t {\bf u} = D \partial_{xx} {\bf u} + {\bf f}({\bf u}) \quad.	
\label{fkpp_d}
\ee
Here ${\bf u} = (u_1,\dots,u_d)$ and ${\bf f} = (f_1,\dots,f_d)$ is such
that ${\bf u}=0$ is an unstable fixed point. 
Then, omitting diffusion, equation $d {\bf u}/ d t = {\bf f}({\bf u})$ can be
chaotic and therefore we expect front solutions connecting the
unstable state ${\bf u}=0$ to the chaotic state to be realizable.

However, as it is well known, there is an easiest way of introducing chaos in
Eq. (\ref{fkpp}), without resting on the generalization to vectorial fields.
Since our interest is in the qualitative effect of chaos in the
reaction terms of (\ref{fkpp}), we can reduce ourselves to studying
the simplest chaotic system, {\em i.e.} $1d$ discrete-time maps.
As we shall see, $1d$ discrete-time maps allow for the analysis of
both chaotic and non-chaotic dynamics, either in the pushed or in the
pulled propagation regimes. Also, they can easily be generalized to
include noise in the system, permitting thereby a comparison 
between deterministic and noisy dynamics. 

Detailed numerical simulations of such systems show that in the
pulled regime the front speed is basically determined by the
unstable fixed point, while the chaotic (noisy) character of the
reaction dynamics plays a limited role. In contrast, in
the pushed regime the presence of chaos (noise) has some
relevance, even if the relationship between chaotic
properties and the front speed appears far from being trivial.
The differences between the chaotic and the noisy situations
are much more evident in the wandering of the front
around its average position. The fluctuations of the
front induced by the chaotic dynamics appear to be bounded on
the examined time scale, while the presence of noise induces
diffusive or sub-diffusive behaviour, in the pushed and pulled
case respectively.

The discrete-time map approach is introduced in
Section II, while the specific map models under study are
introduced in Section III. The numerical results concerning front
speed and front fluctuations are discussed in Section IV, and
finally some concluding remarks are presented in Section V. The details on
the integration algorithm and the estimation of the speed
bounds for the discrete-time map approach are reported in
Appendix A and B, respectively.

\section{Discrete-time map approach}
\label{dtmp}

With respect to Eq. (\ref{fkpp}), there exists a simple way for 
introducing discrete-time maps. This rests on considering the 
case when the reaction term acts in an impulsive way,
\be
f(u) = \Delta t g(u) \sum_n \delta(t-n\Delta t)\quad,
\label{kicked}
\ee
where $g(u)$ is a function whose precise form is not relevant at
this stage, and $\Delta t$ is the period of the forcing.

Consider now a kick given at time $t$. By direct integration of 
Eq. (\ref{fkpp}) between $t$ and $t + \varepsilon$, in the limit 
$\varepsilon \to 0$, we obtain
\be
u(x,t + 0^+) = F(u(x,t))\quad, 
\ee
where 
\be
F(w)=w+\Delta t g(w)\quad.
\ee
$F(v)$ is the reacting map.  In other words we 
have written the evolution of the field between two successive 
kicks in terms of a simple map.

The remaining evolution of the field can be calculated noticing 
that from one kick to the next one, the evolution of the field is
``free'' because the impulsive reactive term by definition does not act. 
Indeed between $t + 0^+$ and $t+\Delta t$, equation 
(\ref{fkpp_d}) reduces to the diffusion equation $\partial_t u 
= D \partial_{xx} u$, whose solution is known
and allows for writing the complete evolution as:
\be
&&u(x,t+\Delta t) \nonumber \\ 
&& \qquad \quad = \frac{1}{\sqrt{4 \pi D \Delta t}}
\int_{-\infty}^{\infty} {\rm d} y \enskip {\rm e}^{-\frac{y^2}{4D \Delta t}}
u(x-y,t+0^+) \nonumber \\  
&& \qquad \quad = \frac{1}{\sqrt{4 \pi D \Delta t}}
\int_{-\infty}^{\infty} {\rm d} y \enskip {\rm e}^{-\frac{y^2}{4D \Delta t}}
F[u(x-y,t)]\quad.  \label{esatta1}
\ee             
Let us notice that the above equation is nothing
but the discrete-time version of the well known
Feynman-Kac formula \cite{freidlin}. A similar approach has been 
recently introduced in \cite{ott} to study the two-dimensional dynamics
of a front separating a chaotic state from a stable steady
one.

We are only left with the evaluation of the convolution 
integral (\ref{esatta1}). Its numerical estimation can be performed 
employing a quite efficient and accurate algorithm recently 
introduced in \cite{torc}. A detailed description of the algorithm is 
given in Appendix A. Once the map is given, then its successive 
iterations account for the whole time evolution of the field 
according to Eq. (\ref{esatta1}). We remark that Eq. (\ref{esatta1}) 
is exact, {\em i.e.} it is not an approximation for small $\Delta t$ 
if Eq. (\ref{kicked}) holds. For the sake of simplicity in the following 
we shall set $\Delta t = 1$.

Of course the evolution of the field will be chaotic or not 
depending on the map. It is well known that the appropriate 
dynamical indicator to discriminate between chaotic 
and non-chaotic dynamics is the maximal Lyapunov exponent $\lambda$, which 
characterizes the divergence in time of nearby orbits. 
This exponent is estimated by 
applying a standard evaluation scheme \cite{benettin} to the
evolution of an infinitesimal perturbation $\delta u (x,t)$ of the
reference orbit in the tangent space
\be
&&\delta u (x,t+1) = \frac{1}{\sqrt{4 \pi D}}
\int_{-\infty}^{\infty} {\rm d} y \enskip {\rm e}^{-\frac{y^2}{4D}} \nonumber \\
&&\qquad \qquad \qquad \qquad \times F^{\prime}[u(x-y,t)] \delta u (x-y,t)
\quad .
\ee
The maximal Lyapunov exponent $\lambda$ is then defined as
\be
\lambda = \lim_{t \to \infty} \frac{1}{t} \ln \frac{||\delta
u(x,t)||}{||\delta u(x,0)||}\quad,
\ee
where $|| \delta u(x,t)||^2 = \int |\delta u(x,t)|^2 dx$,
$\lambda > 0$ representing a signature of chaos. 

\section{The Models}
\label{themodels}

We describe first the two deterministic non-chaotic
maps, which turn out to be useful to illustrate the general approach
and to identify the different regimes present in the system. Starting
from those models two generalization will be proposed, in order to 
assess whether the effects of noise or chaos on the system are relevant. 

\subsection{Deterministic Non Chaotic Models}
\label{determodels}

\begin{figure}[bpt]
\centerline{
\hbox{
\psfig{figure=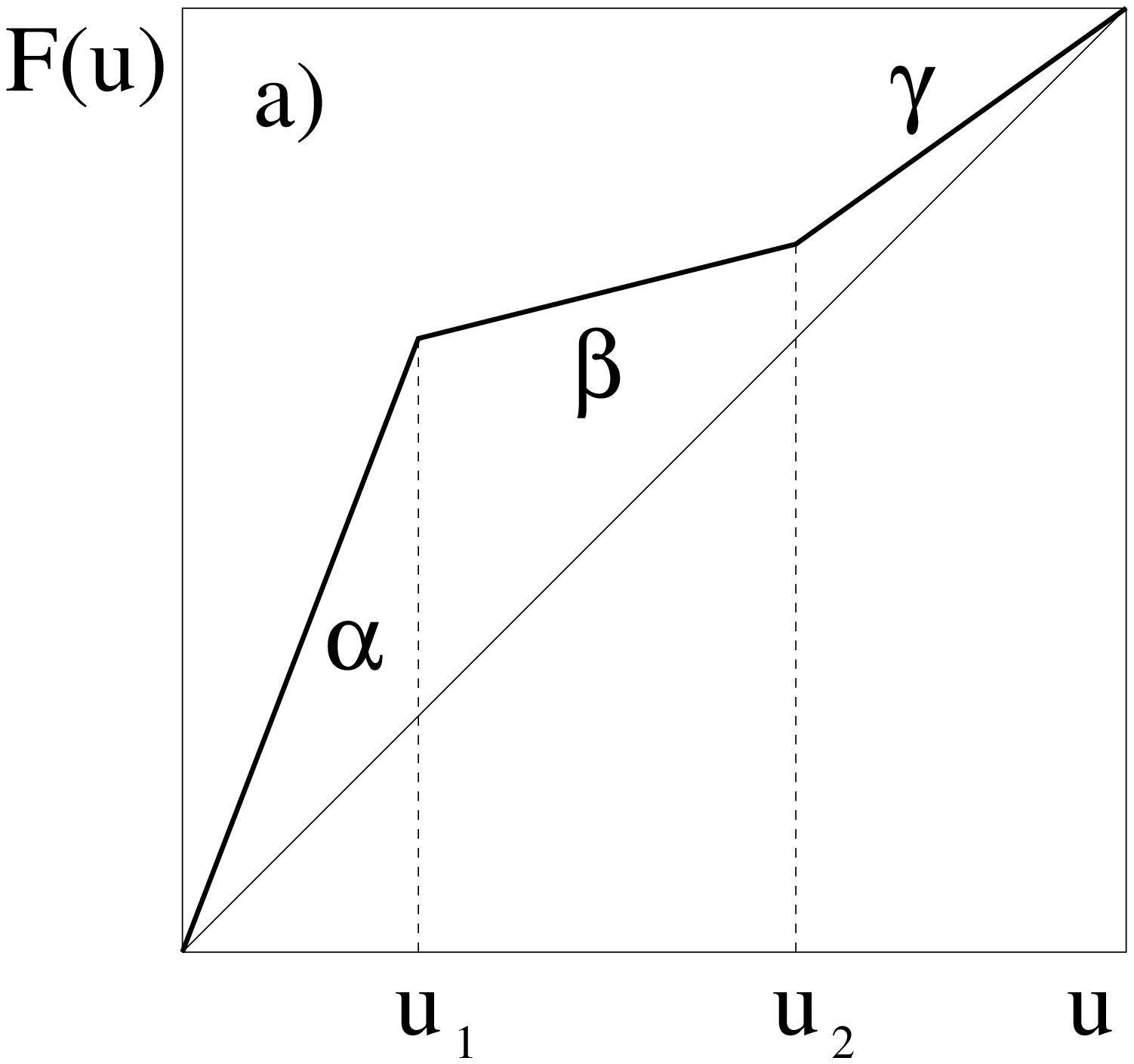,angle=-90,height=4truecm,width=4truecm}
\psfig{figure=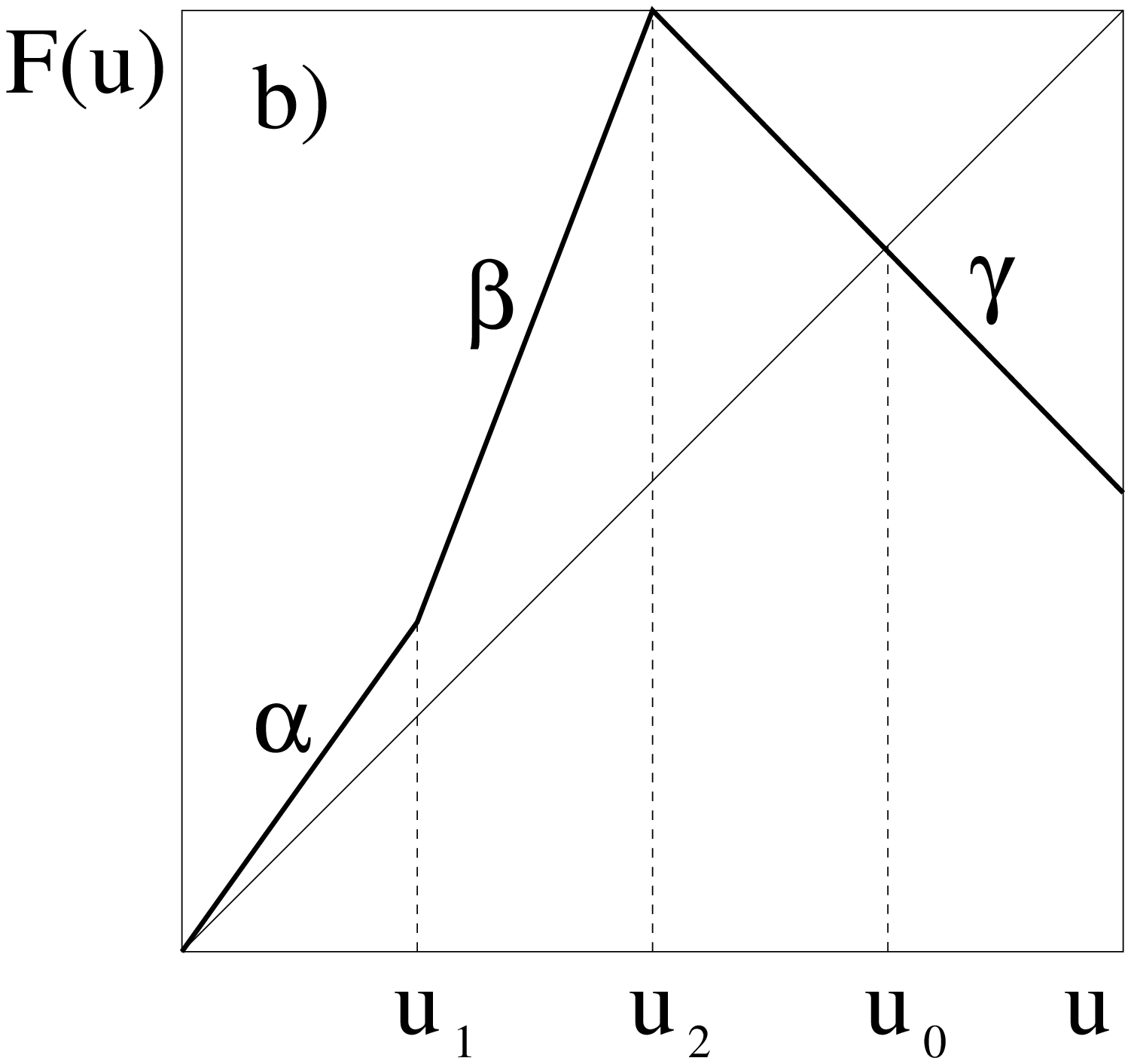,angle=-90,height=4truecm,width=4truecm}
}
}
\caption{Panel a): {\em Map A}. Since $\alpha > \beta$, 
the expected front propagation corresponds to the pulled case 
-- see text. Panel b): {\em Map B}. In this case $\alpha < \beta$
makes it possible to observe pushed dynamics.} 
\label{map_a}
\end{figure}

The first map we introduce has the purpose of reproducing
the usual FKPP behaviour.
As mentioned above, this is given by equation (\ref{fkpp}) where the
function $f$ is chosen in such a way that $u=1$ is a stable fixed
point and $u=0$ is an unstable one.
Inspired by this feature, we propose a deterministic 
non-chaotic map with an unstable fixed point 
in $u=0$ ({\em i.e.} $F(0)=0$ and $|F^{\prime}(0)| > 1$), 
and a stable one $u_0$, 
({\em i.e.} $F(u_0)=u_0$ with $|F^{\prime}(u_0)| < 1$).  
More specifically we define {\em Map A} as  
\be
F(u)=\left\{ 
\begin{array}{cc}
\alpha u, &\qquad 0 \leq u < u_1\\
\beta u + c, &\qquad u_1 \leq u < u_2\\
\gamma u + d, &\qquad u_2 \leq u \leq 1
\end{array}
\right.
\label{map}
\ee
where 
$c=u_1(\alpha-\beta)$, $d=1-\gamma$ and $u_2=(d-c)/(\beta-\gamma)$ 
(see Fig.~\ref{map_a}{\em a}). 

The fixed point $u_0 = 1$ is stable provided that $|\gamma|$ is smaller 
than one. As we shall see, for this system the linear speed is 
$v_{\rm lin} = 2 \sqrt{D \ln \alpha}$ while the slope of the 
exponential part of the leading edge is 
$\mu_{\rm lin} = \sqrt{\ln \alpha /D}$. It is clear that we are always in the pulled
situation if $\alpha \ge \beta$, while if $\beta$ is bigger than
$\alpha$ pushed situations can be observed.

As a second reference model for non-chaotic dynamics we also introduce 
{\em Map B}. Now we set $u_2=1/2$ and $F(u_2)=1$, while $c=1-\beta/2$,
$u_1=c/(\alpha-\beta)$, $d=1-\gamma/2$ and $u_0=d/(1-\gamma)$.
Also in this case we choose $|\gamma| < 1$ in order to have
a stable fixed point in $u_0$. This map is shown in
Fig. \ref{map_a}{\em b}. 
The condition $\alpha < \beta$ allows again 
for pushed situations to be in principle observed.

\subsection{Noisy Models}
\label{noisymodels}

In order to mimic a random environment, in 
\cite{armeropre,rocco1,rocco2} the following noisy RD system 
was considered:
\be
\partial_t u = D \partial_{xx} u + f(u) + g(u) \eta\quad, \label{fn0}
\ee
with
\be
f(u)=u(1-u)(a+u), \qquad g(u)=u(1-u)\quad.  \label{fn}
\ee
The parameter $a$ is a control parameter, by tuning which 
it is possible to change the stability properties of the
invaded state, thereby exploring both pushed and pulled dynamics \cite{armeropre}.
The noise $\eta = \eta(x,t)$ is Gaussian, spatially and temporally 
$\delta$-correlated, and because of the chosen $g(u)$, it does not
affect the fixed points $0$ and $1$.  
The multiplicative noise term present in (\ref{fn0}) 
was first introduced phenomenologically 
as external noise in \cite{armeropre,rocco2}, 
and then rederived in a broader context in \cite{rocco3}. 
Specifically, in \cite{armeropre,rocco2}, 
due to the continuous nature of the model (\ref{fn0}) it was proven
that the appropriate prescription for the evaluation of the noise
term was the Stratonovich one \cite{gardiner}.  

Relevant questions about such systems are typically related 
on one side to the computation of the 
renormalization due to noise of the propagation speed of the front, 
and on the other side to the identification of the general properties of the
wandering of the front around its average position. 

This scenario rests on the decomposition of the propagation 
of the front, solution of (\ref{fn0}), into a global drift 
plus fluctuations. The global drift is associated to an average
deterministic front, where the average is taken over the different
realizations of the noise. This front obeys a deterministic field equation
which can be obtained through a standard procedure from equation (\ref{fn0}) 
(see for example \cite{rocco4}). In the simple case of the choice
(\ref{fn}), this deterministic equation reduces to an equation with
the same reaction term of (\ref{fn}) but with noise-renormalized
parameters, implying thereby a renormalization of the front 
propagation speed \cite{armeroprl}. Notice that this effect is related
to the assumed Stratonovich prescription in the evaluation of the
multiplicative noise term. 

Of course the single 
realizations of the noise induce fluctuations in the shape of the
front, which manifest themselves as fluctuations of the speed around
its average value. This process induces in turn a wandering of the 
front around its average position.
This can be characterized by the root mean square displacement of the
front position $\Delta(t)$.
As a result, the pulled and pushed regime 
differ noticeably. In particular, in the pushed regime the usual diffusive
behavior is proven \cite{armeropre,rocco1} {\em i.e.} $\Delta(t) \sim t^{1/2}$
while in the pulled case a theoretical and numerical analysis shows 
that subdiffusion occurs, with $\Delta(t) \sim t^{1/4}$ \cite{rocco2}.  

Having in mind all this, and starting from the deterministic
non-chaotic models introduced in the previous subsection, namely 
{\em Map A} and {\em Map B}, we introduce the
corresponding noisy maps. Even if a direct comparison between the
continuous model and our discrete-time maps
is beyond the scope of the present paper, 
nevertheless we aim at assessing if the main features
of the continuous FKPP equation with multiplicative noise
are still observable for periodically forced discrete systems.

As a first choice, we consider the stochastic version of 
{\em Map A} reported in Eq. (\ref{map}). We shall refer to this model
as {\em Map NA} and we expect that it will be qualitatively 
equivalent to the noisy FKPP Equation, Eq. (\ref{fn}).
Of course the insertion of noise can be performed on any of the three
segments of the map, identified by their slopes $\alpha$, $\beta$, $\gamma$.
Our strategy will be of inserting noise on one of them, by keeping
fixed the other ones. 
For instance we can consider a situation where 
both $\beta$ and $\gamma$ are fixed while $\alpha$ is chosen 
randomly at any space-time point as
\be
\alpha(x,t) = \alpha_0 + \eta(x,t) \quad.
\ee
Here the noise $\eta$ is distributed according to a flat distribution 
defined in the interval $[-A,A]$, and is $\delta$-correlated 
both in space and time. The model is constructed in such a way 
to maintain $u=0$ and $u=1$ as fixed points. 

The second type of noisy map that we analyze can be introduced 
starting from {\em Map B}. In this case the noise is added to 
the parameter $\gamma$ 
and it has again the same flat bounded distribution of {\em Map NA} 
and the same correlator. In this case, the restriction 
that $|\gamma_0 \pm A|$ be
smaller then one forces the map to have a unique unstable 
fixed point $u=0$ but an infinite number of different stable fixed 
points $u=u_0(\gamma)$. As a matter of fact the interior region of the front, 
developing from the state $u=0$, oscillates stochastically around 
the average fixed point $ \langle u_0 \rangle = d/(1-\gamma_0)$. 
We shall refer to this model as {\em Map NB}. It will be a benchmark
to understand what the effects are of the noise, when it 
affects the saturated part of the front.

\vspace{-0.5cm}

\subsection{Chaotic Models}
\label{chmodels}

\begin{figure}[bpt]
\centerline{
\psfig{figure=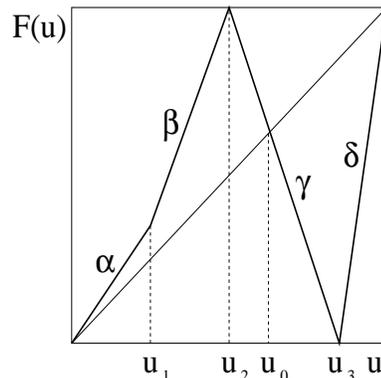,angle=-90,height=5truecm,width=5truecm}
}
\caption{{\em Map CB}. 
A variation of the value of $u_3$ induces
a modification of the corresponding Lyapunov exponent.
}\label{mapcb}
\end{figure}

Modifying {\em Map B} it is possible to introduce a chaotic 
version of the FKPP Equation, which we shall term {\em Map CB}.  
This map is defined by keeping it identical to  
{\em Map B} for $ u \le u_2=1/2$
and modifying the part in the interval $1/2 < u \le 1$ in the following way:
\be
F(u)=\left\{
\begin{array}{cc}
\gamma u + d, &\qquad u_2 \leq u \leq u_3 \\
\delta u + e, &\qquad u_3 \leq u \leq 1 
\end{array}
\right.
\label{map_cc}
\ee                             
where $\gamma=1/(1/2-u_3) < 0$, $\delta=1/(1-u_3) > 0$,
$d=1-\gamma/2$ and $e=1-\delta$, while $u_3$ is arbitrarily chosen
with the restriction to belong to the interval $1/2 < u_3 \le 1$.
{\em Map CB} is represented in Fig. \ref{mapcb}. 

Notice that the fixed points of the map
($u=0$ and $u_0$) are both unstable, because now the value of 
$|\alpha|$ and $|\gamma|$ are bigger than one. With these conditions,
the map appears to be always chaotic, {\em i.e.} the maximal 
Lyapunov exponent $\lambda$ is positive. 

As shown in the next section, in this case 
a localized initial disturbance of the unstable state 
$u=0$ will grow and spread along the system. As a consequence, 
a front will develop connecting the unstable fixed point $u=0$ to a 
chaotic phase, playing the role of the invading state.

\section{Numerical Results and Discussion}
\label{resdisc}

Let us first present some qualitative results concerning the main features 
of deterministic non-chaotic fronts ({\em Map B}), noisy fronts ({\em
Map NB}), and chaotic fronts ({\em Map CB}), in both the pushed and
pulled regimes.

In all the examined cases, the chain length was $L=30,000$, and it 
was initially set everywhere to $0$, with exception 
of a few sites in the center of the 
chain, which were initialized with a disturbance amplitude of ${\cal O} (1)$. 

\begin{figure}[bpt]
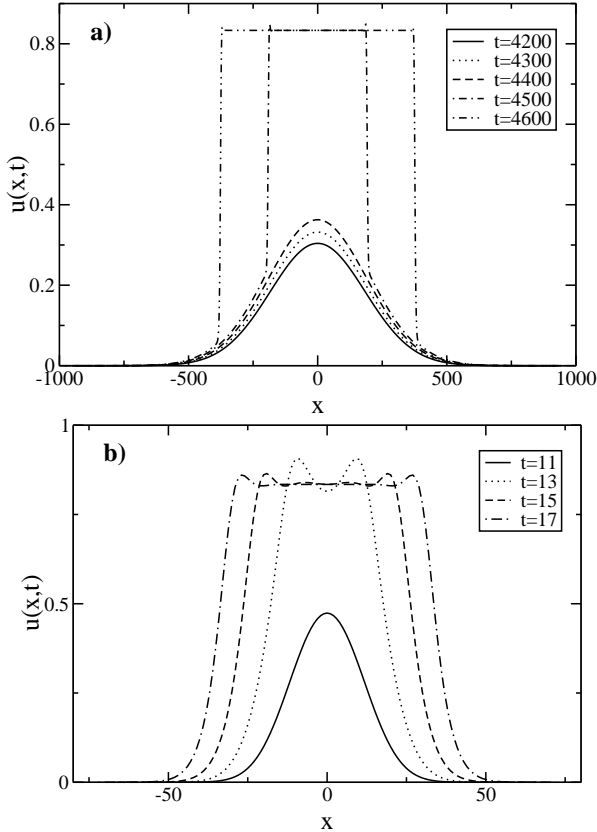

\centerline{\psfig{figure=push_nds.eps,angle=0,height=5.5truecm}}
\centerline{\psfig{figure=pull_nds.eps,angle=0,height=5.5truecm}}
\caption{Front evolution for the FKPP equation with {\em Map B} in
{\em a)} the pushed regime and {\em b)} the pulled one. 
In the pushed regime the values of the parameters are
$\alpha = 1.001$ and $\beta=5.0$, while in the pulled one we set
$\alpha = 1.8$ and $\beta=2.5$. In both cases $D=4$ and $\gamma=-1/2$.}
\label{front_det}
\end{figure}

In Fig. \ref{front_det}{\em a} and \ref{front_det}{\em b} we show the
deterministic non-chaotic behaviour of pushed and pulled fronts 
respectively. The corresponding dynamics is selected by changing the 
relative weight of the parameters $\alpha$ and $\beta$, as explained
in Section \ref{determodels}. The realization of the pushed or pulled 
regime is checked by direct measurement
of the front speed (see the precise definition in the next Section). 
The linear speed corresponding to case shown in Fig. \ref{front_det}{\em a} 
is $v_{\rm lin} = 0.126$, while the measured value is
$v_{\rm f} = 1.444$, confirming thereby the realization of the pushed
regime. As for Fig. \ref{front_det}{\em b}, we measure 
$v_{\rm f} = v_{\rm lin}$, with $v_{\rm lin} = 3.066$, corresponding
indeed to pulled dynamics.

In the pushed case it is evident an abrupt jump from an 
initial evolution where the dynamics
is ruled by the linear mechanisms (characterized by a Gaussian
shaped perturbation propagating with velocity $v_{\rm lin}$) 
to a situation where nonlinearities set
in and the front saturates in its central part and starts
propagating with a velocity $v_{\rm f} > v_{\rm lin}$. 
In contrast, in the pulled situation (depicted in Fig. \ref{front_det}{\em b})
the transition from the initial Gaussian perturbation to a saturated 
propagating front is smoother, since now
the only mechanism responsible for propagation is the linear one and
the speed does not increase above the linear value.

\begin{figure}[bpt]
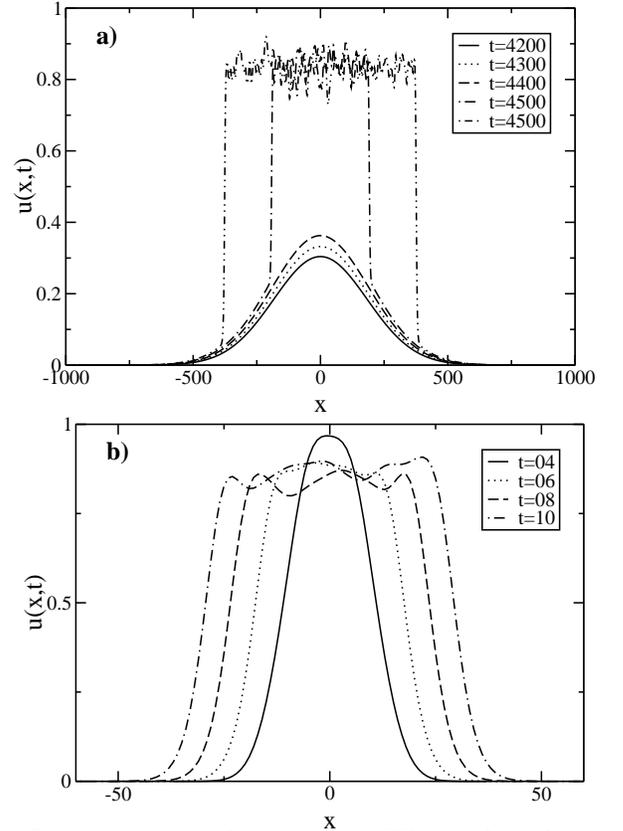

\centerline{\psfig{figure=push_nnn.eps,angle=0,height=5.5truecm}}
\centerline{\psfig{figure=pull_nnn.eps,angle=0,height=5.5truecm}}
\caption{Front evolution for the FKPP equation with {\em Map NB} in
{\em a)} the pushed regime and {\em b)} the pulled one. 
In the pushed regime the values of the parameters are $D=4$,
$\alpha = 1.001$ and $\beta=5.0$, while in the pulled one we set
$D=4$, $\alpha = 1.8$ and $\beta=2.5$. The noise is set on the parameter 
$\gamma$ with $A=0.45$, with $\gamma_0 = -1/2$.}
\label{front_noisy}
\end{figure}

Noisy front propagations as given by {\em Map NB} are shown in Fig. 
\ref{front_noisy}{\em a} and \ref{front_noisy}{\em b} for 
pushed and pulled dynamics 
respectively. In this case the values of the parameters are the same
as in Fig. \ref{front_det}, with the only exception of the $\gamma$
parameter, which is the one affected by the noise. 

It is remarkable the fact that there seems to be no effect of noise 
on the value of the speed. As a matter of fact, the measured speeds have the
values $v_{\rm f} = 1.444$ (against the linear value $v_{\rm lin} = 0.126$) 
in the pushed case and again $v_{\rm f} = v_{\rm lin} = 3.066$ in the pulled
one. These values coincide with the ones obtained in the
deterministic non-chaotic dynamics.
However, by analogy with the continuous model, we could naively expect a
renormalization of the front speed to occur in both pulled and 
pushed regimes. The fact that this does not seem to be the case 
is related to the intrinsic discreteness of our models. 
In the next Section we shall analyze this point in greater detail. 

\begin{figure}[bpt]
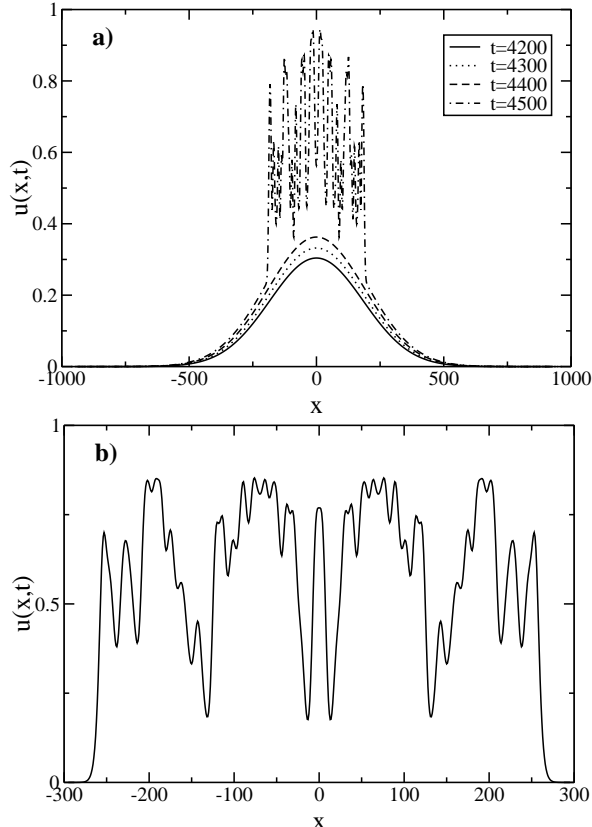

\centerline{\psfig{figure=push_dsk.eps,angle=0,height=5.5truecm}}
\centerline{\psfig{figure=pull_dsk.eps,angle=0,height=5.5truecm}}
\caption{Front evolution for the FKPP equation with {\em Map CB} with
$u_3 =1$ in
{\em a)} the pushed regime and {\em b)} the pulled one. The parameters
$\alpha$, $\beta$,and $D$ are again chosen as in the previous figures,
while the parameter $\gamma$ is now set to the value $\gamma=-2$. The
snapshot shown in Panel {\em b} is taken at time $t=90$.}
\label{front_ch}
\end{figure}

Our results for the chaotic models are shown in Fig. \ref{front_ch}, again for
both pushed and pulled dynamics. 
As it is evident by comparison of Fig. \ref{front_noisy} and
\ref{front_ch}, chaos and noise seem to affect in a qualitatively
equivalent way the front structure, at least when the noise is set on
the $\gamma$ parameter. This could suggest the possibility of easily 
building up an effective noisy model for a general underlying
chaotic dynamics. However as we shall see in the next Section, an effective 
equivalence between the two models is far from being trivial, due to the
different role played by noise and chaos on the renormalization
properties of the speed of the front. This is clear in the pushed
regime, where the front speed takes the value $v_{\rm f} = 1.276$, 
smaller than in the corresponding non-chaotic and noisy models,
indicating a deep difference between the respective dynamics. 

Finally notice that all the fronts shown in the previous figures
manifest the same dynamical evolution at short times up 
to $t^* = {\cal O}(1/\ln(\alpha))$. 
As a matter of fact, in the pushed case the front evolution associated 
to the three maps become distinguishable at
times of the order of $t=4,400$, with $\alpha=1.001$, while
for the fronts in the pulled regime
the evolution of the three maps are practically identical up 
to time $t=12$, since now the $\alpha$ parameter is much bigger
(namely, $1.8$). Afterwards it becomes possible to distinguish 
the different evolutions in both situations.

In order to investigate at a more quantitative level 
the different dynamical behaviour 
of fronts in presence of noisy or chaotic dynamics, we analyze now
the front propagation 
speed and the fluctuations of the front position. 

\subsection{Speed of the Front}
\label{velfront}

It is not difficult to show that for 
the discrete-time map version of the FKPP Equation,
the propagation speed $v_{\rm f}$ is always bounded in the interval
$[v_{\rm lin},v_{\rm max}]$, where
\be
v_{\rm lin} = 2 \sqrt{D \ln(F^{\prime}(0))} 
\label{vmin2}
\ee
and
\be
v_{\rm max} = 2 \sqrt{D \ln \left[\max_{0 \le u \le 1} \, \frac{F(u)}{u}\right]}\quad.
\label{vmax2}
\ee
Justification of (\ref{vmin2}) and (\ref{vmax2}) is given in Appendix B. 
As mentioned before, if $v_{\rm f} = v_{\rm lin}$, then the dynamics is
pulled, while if $v_{\rm f} > v_{\rm lin}$ the corresponding dynamics will
be pushed. 

From the numerical point of view, the measurement of the speed has been
performed in the following way.
After having initialized to zero all the chain apart a localized disturbance
$u(x,0) = {\cal O}(1)$ for $-\xi \le x \le \xi$,
at each time step the rightmost $r(t)$
and the leftmost position $l(t)$ of the front are considered,
\be
&&r(t)= \max \{x \; | \; u(x,t) > \vartheta \}\quad, \\
&&l(t) = \min \{x \; | \; u(x,t) > \vartheta \}\quad,  
\ee
where $\vartheta$ is a preassigned threshold. The position of the front
is simply $x_f^r(t)=r(t)-\xi$ or $x_f^l(t)=\xi-l(t)$, depending on which of 
the two moving fronts is considered, and the speed results in 
\be
v_{\rm f} = \lim_{t \to \infty} \frac{x_f^r(t)-x_f^l(t)}{2t} \quad .
\label{vel}
\ee
We have verified that $v_{\rm f}$ does not depend on the chosen threshold
and obviously the time limit reported in (\ref{vel}) should be taken
after the limit $L \to \infty$.  

\begin{figure}[bpt]
\centerline{
\psfig{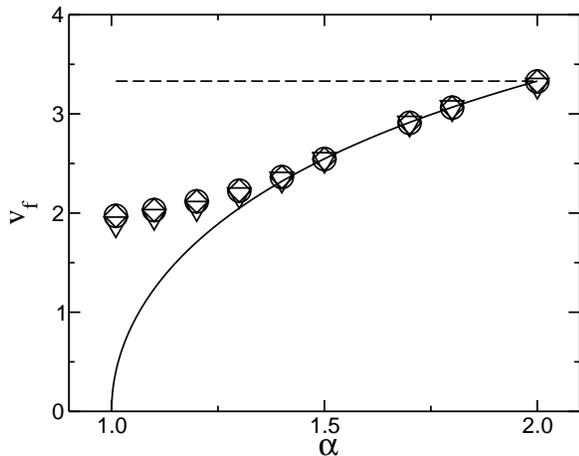}}
\caption{Front speeds $v_{\rm f}$ as a function of the parameter $\alpha$
for the {\em Map CB} ($u_3 = 1$) (triangles), {\em B} (circles) 
and {\em NB} (diamonds) with $\beta=2.5$ and $D=4$.
The dashed line indicates $v_{\rm max} = 2
\sqrt{D \ln 2}$ and the solid one $v_{\rm lin} = 2 \sqrt{D \ln \alpha}$.
The length of the chain is $L=10,000$, the front has
been followed for a time $t=1,600$, and the average is
performed over $400$ different initial conditions. 
}
\label{vel_cb}
\end{figure}

We present now a more detailed investigation of the dependence 
of the speed on both noise and chaos. Consider first Fig. \ref{vel_cb}.
Our main result is that the ``structural''
features of the map (its concave or convex character) are 
sufficient to provide information about the type of
front propagation observable and the chaotic, noisy or non-chaotic
nature of the dynamics seems to be secondary. This conclusion relies
on the measurement of the front speed
for {\em Map B}, {\em Map NB}, and {\em Map CB} with $u_3 = 1$. 
We have performed the simulation by keeping 
$\beta=2.5$ and $\gamma=-1/2$ fixed, and changing the   
$\alpha$ values in the interval $1 \le \alpha \le 2$. 
As shown in Fig. \ref{vel_cb}     
$v_{\rm f} > v_{\rm lin}$ for $\alpha <  1.4$, and therefore for
$\alpha > 1.4$ the linear mechanism prevails on the nonlinear
one and the front is pulled from the linear instability of
the leading edge. The same choice of parameters has been done also
for maps {\em CB} ($u_3 = 1$) and {\em NB} in such a way that the 
three maps essentially coincide for $u < u_2=1/2$. 

As we can observe from Fig. \ref{vel_cb} the speeds of the fronts are
almost identical in the three cases. Only in the strongly pushed
regime, that is for values of $\alpha$ very close to 1,
 the propagation for the chaotic model appears to be slightly slower than 
in the other models, but the discrepancy is indeed very small.
For values of $\alpha$ as small as $\alpha = 1.001$ 
the observed values for the velocities obtained for  
both {\em Map B} and {\em NB} are $v_{\rm f}=1.965$, while for {\em Map CB}
(with $u_3=1$)
we obtain $v_{\rm f}=1.884$, corresponding to a discrepancy of the order of 4\%. 

In order to better analyze these findings, we investigate the
dependence of the speed separately as a function 
of the intensity of the noise $A$ and of the Lyapunov coefficient 
$\lambda$. 

As far as the effect of noise is concerned, 
we consider first {\em Map NB}. We set noise on the three
parameters of the map, {\em i.e.} $\alpha$, $\beta$ and $\gamma$, 
but the corresponding changes of the speed appear basically 
negligible in all cases. 

\begin{figure}[bpt]
\psfig{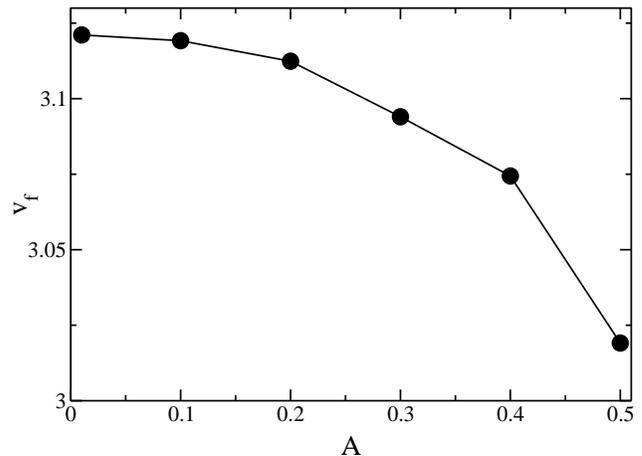}
\caption{Dependence of the front speed $v_{\rm f}$ on the intensity of noise
$A$ in {\em Map NB}. Here the noise is inserted on the parameter 
$\gamma$, with $\gamma_0 = -1/2$. The values of the other parameters
are $\alpha = 1.001$ and $\beta = 5$ (strongly pushed), and $D=4$.}
\label{qaccd}
\end{figure}

More interesting is the case corresponding to {\em Map NA}. In this
situation, there is still no effect of noise in the pulled regime,
while in the pushed case we find a decrease
of the speed with the noise amplitude, for noise added to any of
the three parameters. Anyway, the strongest decrease
has been observed when the noise affects the parameter $\gamma$. 
This case is the one reported in Fig. \ref{qaccd}. 
At this point it is worth to remark that the very weak 
velocity renormalization due to noise observed in the present
context is not in contradiction with the results reported in
\cite{armeroprl}, where a noticeable variation of the
front speed with the noise amplitude was found. 

This is so because the intrinsic discrete nature of our 
system does not leave room to any
ambiguity in the definition of the multiplicative noise term (the
so-called {\em Ito-Stratonovich dilemma} \cite{gardiner}).
As a matter of fact, in the limit $\Delta t \to 0$
our model would correspond to a continuous model with multiplicative 
noise inserted according to the Ito's prescription. 

\begin{figure}[bpt]
\centerline{
\psfig{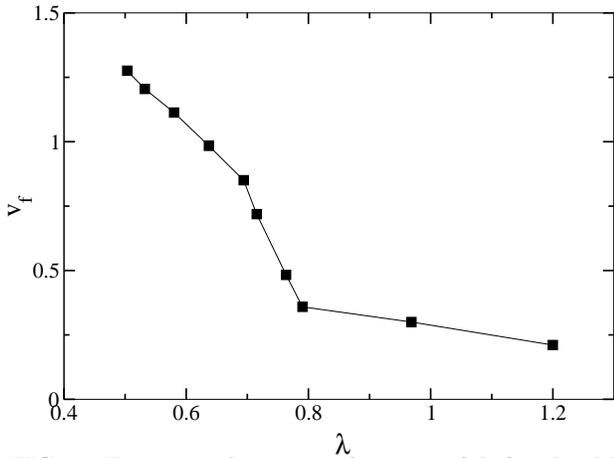}
}
\caption{Front speeds $v_{\rm f}$ as a function of $\lambda$ for the {\em Map CB}.
The values of the parameters are in this case
$D=4$, $\alpha = 1.001$, $\beta=5$, $u_1 = 0.3751$. 
The length of the chain is fixed for all the simulations to 
$L=30,000$, the front has
been followed for a time $t=5,000$, and the average are
performed typically over $1,000$ different initial conditions.
}
\label{lya_001}
\end{figure}

The second point worth of investigation is the effect of chaos.
In this case we performed a measurement of the speed of the front as a
function of the Lyapunov coefficient in the strongly pushed regime.
A different degree of chaoticity is obtained by tuning the value of $u_3$ 
in {\em Map CB}, and can be evaluated via the corresponding 
Lyapunov exponent $\lambda$. Our results are plotted in Fig. \ref{lya_001}. 

The speed exhibits a strong decrease as a function of $\lambda$, which
seems to indicate a behaviour similar to noisy case. However, 
by direct comparison of Figs. \ref{qaccd} and \ref{lya_001}, it appears 
evident that chaos affects the system in a quite more relevant way 
then noise. 

\subsection{Fluctuations of the front}
\label{flucfront}

Now we consider the root mean square displacement of the front position
$x_f(t)$,
\be
\Delta (t) = \sqrt{\langle x_f^2(t)\rangle - \langle x_f(t) \rangle^2}
\quad,
\label{std}
\ee                     
where the average is taken over different initial conditions
for the chaotic case and over many distinct noise realizations for
the noisy case. 
Different scalings for $\Delta(t)$ are observed 
depending on the type of map and on how the noise enters the dynamics.

Let us examine first {\em Map NA}. 
We have studied the three different cases corresponding on applying
noise on $\alpha$, $\beta$ or $\gamma$.
In the pulled case the noise does not induce any wandering
of the front when it is added to $\beta$ or $\gamma$:
In such cases $\Delta(t) \to const.$ in the limit $t \to \infty$.
In contrast, if the noise is applied to the $\alpha$
parameter, sub-diffusive behavior is indeed observed.
These results can be justified recalling that in the pulled
case the dynamics of the front is determined 
in the leading edge. This region corresponds 
to small $u$ values, and therefore only a (stochastic) change in
$\alpha$ is expected to affect the behavior of the system.

\begin{figure}[bpt]
\centerline{
\psfig{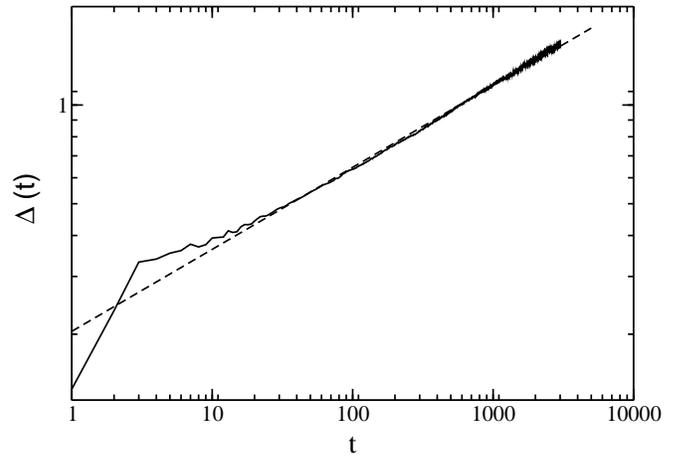}
}
\caption{Fluctuations $\Delta(t)$
as a function of time in a log-log scale for {\em Map NA},
with noise on the $\alpha$-parameter.
Namely $\alpha_0=10$, $A=4$, $\beta=1.0$, $\gamma=-0.2$, and $D=4$.
The data have been averaged over $4,400$ different 
noise realizations for a chain of length $L=60,000$,
with bounded noise. 
The dashed line refers to a power law $t^q$ with exponent $q = 1/4$. 
In this case the linear speed is $v_{\rm lin} = 6.0697$, while the
measured one is $v_{\rm f} \simeq 6.069$: we are clearly in the pulled case.
}
\label{s1e4}
\end{figure}

\vspace{-0.4cm}

\begin{figure}[bpt]
\centerline{
\psfig{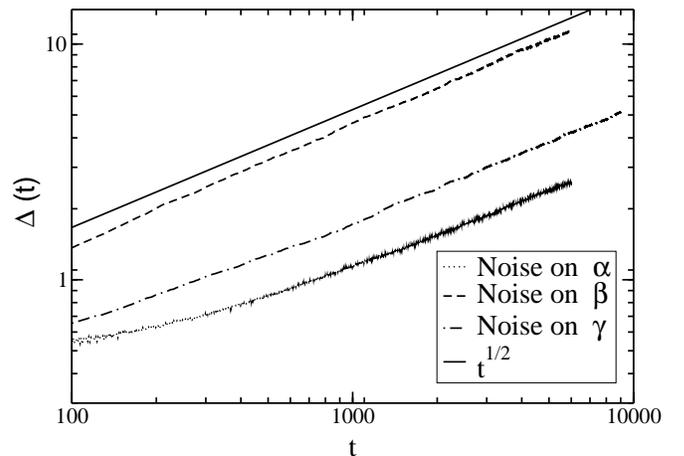}
}
\caption{Fluctuations $\Delta(t)$
as a function of $t$ in a log-log scale for {\em Map NA},
with noise added to the parameter $\alpha$
(with values of the 
parameters $\alpha_0=2$, $A=1$, $\beta=6$, and
$\gamma = -0.2$),
$\beta$ (with values of the 
parameters $\alpha=2$, $\beta_0=6$, $A=3$ and
$\gamma = -0.2$) or
$\gamma$ (with values of the
parameters $\alpha=1.1$, $\beta=2.5$,
$\gamma_0 = -0.5$, and $A=0.4$); always $D=4$. 
The solid line represents the power law $t^q$ with $q=1/2$.
We are in the pushed situation since
$v_{\rm f} > v_{\rm lin}$ for all three examined cases.
Namely for noise acting on $\alpha$, $v_{\rm f}=4.436$
and $v_{\rm lin}=3.330$; for noise acting on $\beta$ 
the linear speed is the same while $v_{\rm f}=4.406$; 
for noise acting on $\gamma$, $v_{\rm lin}=1.235$,
while $v_{\rm f}=3.075$.
}
\label{s1e2}
\end{figure}

On the other hand, in the pushed regime we observe that in all
three cases the noise has the same effect on the front
wandering and it leads to a diffusive
behaviour for the front positions (see Fig. \ref{s1e2}). 
This effect is also reasonable since now
the front propagation is also related to the regions 
where the field takes values of ${\cal O}(1)$, 
and not only to the leading edge. Therefore the effect of 
the noise will be equally relevant, independently of what parameter 
it is applied to.

To complete the analysis related to the noisy dynamics
we have also considered {\em Map NB} with noise
on the parameter $\gamma$ only. In the pulled case,
similarly to the {\em Map NA} case, we observe saturation of
the root mean square displacement $\Delta(t)$. In the pushed case
$\Delta (t)$ indeed grows in time, but within
our time window and for the examined parameters
($\alpha=1.9$, $\beta=2.5$, $\gamma_0=-0.5$
and $A=0.4$, with $v_{\rm f}=v_{\rm lin}=3.2046$),
we do not observe any clear scaling.

\begin{figure}[bpt]
\psfig{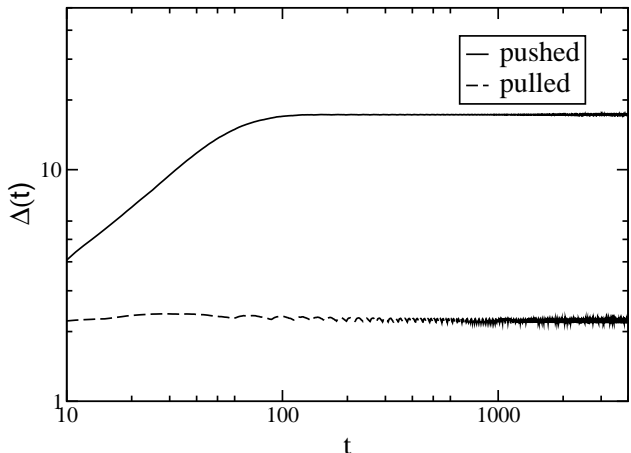}
\caption{Root mean square displacement $\Delta(t)$  as a function of time for {\em
 Map CB} with $u_3=1$. The values of the parameters are in this case
$\alpha = 1.1$ (pushed) and $\alpha = 1.8$ (pulled), $\gamma=2.5$, $\beta=-2$, 
and $D=4$.
The length of the chain was $L=60,000$ and the average was performed
over 2,000 initial conditions.} 
\label{rmsdch}
\end{figure}

Finally we address the chaotic case. Our results are reported in
Fig. \ref{rmsdch}. 
In this situation, in contrast with the stochastic case, we observe
for the pulled as well as for the pushed regime that 
$\Delta(t)$ saturates to a constant  values.
We have verified this for {\em Map CB} with $u_3 = 1$
(see Fig. \ref{rmsdch}). An analogous result
has been observed by Rudzick {\it et al.} \cite{pikov}
for a Coupled Map Lattice (CML) model. In that case
a diffusive behaviour for the wandering of the front position 
is observed for small diffusive coupling, while
for sufficiently strong coupling $\Delta(t)$ is shown to saturate.
The authors of \cite{pikov} argue that this result can be 
explained as follows. For small couplings, the spatial 
correlation between adjacent sites is
negligible on a distance corresponding to their lattice spacing.
Therefore, the successive positions occupied by
the front in its time evolution will be completely decorrelated,
and the front dynamics induced by the local chaotic
evolution can be described in term of a stochastic process.
In contrast, when the coupling becomes large enough, 
the lattice sites become correlated on a length
scale larger than the lattice spacing, and this implies that
a description of the front evolution 
in terms of a stochastic process in not appropriate
any longer, not even for chaotic maps. 
Notice that the CML model is a spatially discrete model, which can be
considered a fair approximation of a continuous space system 
only for sufficiently large
diffusive couplings. Therefore in our case we always expect that for chaotic 
maps $\Delta(t)$ cannot grow indefinitely in time, and saturation is
the only behavior observable. 
This indicates that the chaotic evolution cannot be simply
reduced to some erratic behaviour sharing
common dynamical features with the noisy systems.

\section{Concluding Remarks}
\label{concl}
In this paper we have studied the front propagation
in reaction diffusion systems with a periodically 
forced reaction term. For this case we have been able to rewrite
the evolution of the system as a convolution of a spatial
propagator and a discrete-time map. By implementing a quite
efficient algorithm for the evaluation of the
convolution integral, we have numerically studied  
the front dynamics for deterministic non-chaotic and chaotic
maps as well as for noisy maps. An analogy with the usual
FKPP problem can be drawn for non-chaotic maps, allowing
to obtain the expression for the lower and upper bound
for the speed of the front. Moreover,
even when the reaction dynamics presents an unstable fixed point 
coexisting with a chaotic (or noisy) behaviour, the
analogy with the FKPP problem still holds, once 
the chaotic (noisy) phase is identified with the stable
fixed point of the usual FKPP reaction term.
In the pulled regime the presence of chaos (noise) plays a poor
role and the front speed $v_{\rm f}$ is essentially determined by
the dynamical evolution around the unstable fixed point.
On the contrary in the pushed case $v_{\rm f}$ depends in a nontrivial
way on the details of the chaotic (noisy) behaviour.
Unfortunately, the relationship between the chaotic properties
({\em e.g.} the Lyapunov exponent) and the values of the observed
$v_{\rm f}$ does not appear simple.
In particular, the effect of chaos seems to be much stronger than 
that related to the noise: 
{\em e.g.} $v_{\rm f}$ changes noticeably for chaotic reaction
dynamics at varying the Lyapunov exponent, while it remains
almost constant as a function of the strength of the noise.
In contrast, the noise induces a wandering of the
front around its average position, which is diffusive
in the pushed case and sub-diffusive in the pulled one.
Our results confirm the same scaling laws found in 
continuous models with multiplicative noise 
\cite{armeropre,rocco1,rocco2} for the time evolution of the 
mean square displacement $\Delta(t)$ associated to 
the front position fluctuations, namely $\Delta(t) \sim t^{1/2}$
in the pushed case and $\Delta(t) \sim t^{1/4}$ in the
pulled situation. This suggests that such results are universal
in that they do not depend on the details of the system.
When a chaotic dynamics is considered both in the pushed and 
in the pulled case a saturation of $\Delta(t)$ to a constant 
value is observed, and this is consistent with previous results 
obtained for coupled map lattice with strong diffusive 
couplings \cite{pikov}.

\acknowledgments
We acknowledge useful discussions with R. Kapral and W. van
Saarloos. A.V. thanks the University of Alaska Fairbanks for the warm
hospitality during the final write up of this article.

\begin{appendix} 

\section{Details on the integration algorithm}

Let us now present the numerical details concerning
the integration of Eq. (\ref{esatta1}), 
\be
&&u(x,t+\Delta t) = \nonumber \\
&&\qquad\frac{1}{\sqrt{4 \pi D \Delta t}}
\int_{-\infty}^{\infty} {\rm d} y \enskip 
{\rm e}^{-\frac{y^2}{4D\Delta t}} F[u(x-y,t)]\quad,
\label{esattaa}
\ee                                                              
which represents the convolution between the field at time $t+\varepsilon$
({\em i.e.} $u(x,t+\varepsilon)=F[u(x,t)]$) and the Gaussian Kernel
\be
K(x) = \frac{1}{\sqrt{4 \pi D\Delta t}}  
\enskip {\rm e}^{-\frac{x^2}{4D\Delta t}}\quad.
\label{kernel}
\ee
The integration (\ref{esattaa}) can be performed 
on a computer once the field is discretized on a grid
of resolution $\Delta x$. The integral 
reported in Eq. (\ref{esattaa}) reduces then to the sum
\be
&&u(i \Delta x,t+\Delta t) = \nonumber \\
&&\qquad\qquad\sum_{m=-N/2}^{N/2} 
K(m \Delta x) F[u((i-m) \Delta x,t)]\quad,
\label{discrete}
\ee          
where the system length is $L = N \Delta x$ and periodic
boundary conditions are assumed along the chain.

In order to improve the integration speed we have restricted the
sum (\ref{discrete}) to a small number $M$ of neighbours of the
site $i$ and to ensure a good precision the true kernel
has been substituted by a ``modified'' kernel $C(m)$, 
chosen to minimize the integration error. With these choices
Eq. (\ref{discrete}) now reads:
\be
&&u(i\Delta x,t+\Delta t) = \nonumber \\
&&\qquad\qquad\sum_{m=-M}^{M} 
C(m) u((i-m) \Delta x,t+\varepsilon) \quad.
\label{modified}
\ee      
The problem is now to determine the coefficients
$\{ C(m) \}$ for $m=-M, \dots, M$. In order to do this,
we first suppose that an appropriate Fourier basis 
made up of $(2 M +1)$ elements, able to
well approximate the field on the chosen grid,
can be written as
$$
\left\{
\exp[j (k \alpha) x]
\right\}\quad,
$$
where the parameter $\alpha$ will be determined later.
By rewriting Eq. (\ref{modified}) on this basis
one is left with the following set of equations:
\be
&&\sum_{m=-M}^{M} C(m) \exp[-j (k \alpha) m \Delta x] \nonumber \\
&& \qquad \qquad \qquad \qquad = \exp[-(k \alpha)^2 D \Delta t]\quad. 
\label{set} 
\ee      
Due to the Kernel symmetry we can reduce the system
of $(2 M+1)$ equations (\ref{set}) to a set of $M$ equations.
This because the elements of the Kernel are
symmetric around $m=0$ (i.e. $C(m)=C(-m)$), and are related by
the normalization condition
$$
C(0) = 1 - \sum_{m \ne 0} C(m) \quad .
$$
We can determine the $M$ independent 
elements of the Kernel $C(m)$ by solving the system (\ref{set})
as a function of the parameter $\alpha$, once the integration
time step $\Delta t$, the diffusion coefficient $D$, and the
spatial resolution $\Delta x$ are fixed. The optimal 
choice of the parameter $\alpha$ is achieved by requiring that
the first six cumulants of the discretized Kernel reproduce
those of the true Kernel within a precision of one
part over a million and that the quadratic sum
$ \sum_{m=-M}^M C^2(m)$ is minimal.
In the present paper, we have always used $D=4$, $\Delta t =
\Delta x = 1$ with $M=15$ and a parameter value
$\alpha = 0.13$.

This integration scheme has been previously introduced as a
possible alternative to the usual pseudo-spectral codes
to evaluate the dynamical evolution of the complex 
Ginzburg-Landau equation
and of the Fitzhugh-Nagumo equation in one and two dimensions
\cite{torc}.

\section{The speed bounds for discrete time maps}
The linear speed (\ref{vmin2}) for discrete maps
can be obtained with simple consideration just following
the standard reasoning used for the derivation of $v_{\rm lin}$ in the
continuous time limit:
\be
   \frac{\partial}{\partial t} u =
   D \nabla^2 u  + f(u)\quad.
   \label{eq:rdapp}
\ee

As we have already mentioned, the leading edge of the
propagating front has an exponential shape:
\be
   u (x,t) \simeq e^{-\mu x + \lambda(\mu) t}\quad.
   \label{eq:espshapeapp}
\ee

Inserting (\ref{eq:espshapeapp}) in (\ref{eq:rdapp})
and linearizing around $u = 0$ one obtains:
\be
   \lambda(\mu) = D \mu^2 + f'(0)\quad.
   \label{eq:relabapp}
\ee
A stationary phase argument gives a selection criterion
which allows for the determination of the front speed as
\be
   v_{\rm lin} = \min_{\mu} \frac{\lambda(\mu)}{\mu} = 2 \sqrt{{D} f'(0)}\quad.
   \label{eq:vf}
\ee
Let us consider now the discrete-time reaction case, 
\be
   \frac{\partial}{\partial t} u = D \nabla^2 u + 
   \sum_{n=-\infty}^\infty g(u) \delta(t - n)\quad,
   \label{eq:raddiscr}
\ee
where we have adopted $\Delta t=1$.
Indicating with $F(u)$ the reacting map one gets:
$$u(x,t+0^+) = F(u(x,t))\quad,$$
and integrating the diffusion equation $\partial_t u=D \nabla^2 u$
between $t+0^+$ and $t+1$ one obtains:
\be
   u(x, t+1) = \frac{1}{\sqrt{4 \pi D}}
       \int e^{-\frac{w^2}{4D}} F \left ( 
       u(x - w, t) \right ) {\mathrm d} w\quad,
   \label{eq:fwykacdimpmapapp}
\ee
Assuming the shape (\ref{eq:espshapeapp}) and linearizing around
$u = 0$, {\em i.e.}: $F(u) \simeq F'(0) u$, a simple
Gaussian integration gives:
$$ e^{\lambda(\mu) (t+1)- \mu x} 
\simeq e^{\ln F'(0) + D \mu^2-\mu x+ \lambda(\mu) t}\quad.$$
The above result implies
$$ \lambda(\mu) = \ln F'(0) + D \mu^2\quad,$$
which is nothing but Eq.~(\ref{eq:relabapp}) now with $\ln F'(0)$
in place of $f'(0)$. The same selection criterion
gives
\be
   v_{\rm lin} = 2 \sqrt{D \ln F'(0)}\quad.
   \label{eq:vfdiscr}
\ee

In order to estimate $v_{\rm max}$ in the discrete case,
let us consider again the continuous equation
(\ref{eq:rdapp}) with the shape of $u(x,t)$ given
by (\ref{eq:espshapeapp}). It is straightforward to show that
\be
   \lambda(\mu) \le  D \mu^2 + \max_{0 \le u \le 1} \frac{f(u)}{u}\quad.
   \label{dis_con}
\ee
From the above inequality it can be shown that
\be
v_{\rm max} = 2 \sqrt{D \max_{0 \le u \le 1} \, \frac{f(u)}{u}}\quad.
\label{vmax_2}
\ee
In the discrete case we write the following inequality
\be
   &&u(x, t+1) \le \frac{1}{\sqrt{4 \pi D}}
       \left( \max_{0 \le u \le 1} \frac{F(u)}{u} \right)\nonumber \\
   &&\qquad \qquad \qquad \qquad \times \int e^{-\frac{w^2}{4D}} 
       u(x - w, t) {\mathrm d} w\quad,
   \label{dis}
\ee
and assuming that the leading edge has the shape reported
in (\ref{eq:espshapeapp}) one obtains
\be
\lambda(\mu) \le D \mu^2 + \ln\left[\max_{0 \le u \le 1} \frac{F(u)}{u}\right]\quad.
   \label{dis2}
\ee
This equation is analogous to (\ref{dis_con})
therefore in the discrete case the upper bound for the
speed is now 
\be
v_{\rm max} = 2 \sqrt{D \ln \left[\max_{0 \le u \le 1} \, \frac{F(u)}{u}\right]}\quad.
\label{vmax_3}
\ee

\end{appendix}

\end{multicols}

\end{document}